%%%%Phyzzx file
\line{ \hfill ULB-TH 15/94}   \line{ \hfill August 1994}
\bigskip\bigskip\bigskip\bigskip\bigskip \centerline  {\bf THE BLACK
HOLE HISTORY IN TAMED VACUUM }  \bigskip\bigskip\bigskip
\centerline{F. ENGLERT \foot{E-mail: fenglert at ulb.ac.be}}
 \centerline {\it Service de Physique Th\'eorique} \centerline{\it
Universit\'e Libre de Bruxelles, Campus Plaine, C.P.225 }
\centerline{\it Boulevard du Triomphe, B-1050 Bruxelles, Belgium}
\bigskip\bigskip  \noindent {\bf Abstract} \medskip Quantum physics at
scales large compared to the Planck scale is described in the
framework of classical space-time geometries. A criterion for
selecting these backgrounds out of quantized gravity is proposed.  It
leads to an instability of the  black-hole geometry, as experienced by
motion across the horizon, against emission of Hawking quanta.  A
phenomenological treatment of the evaporation process perceived by
external observers who do not cross the event  horizon is presented.
Evaporation  occurs within a topologically trivial ``achronon"
geometrical background devoid of horizons and singularities describing
a collapse frozen up to decay time scales. It is ignited as in the
conventional theory from pair creation out of the vacuum of the
collapsing star of mass $M$, but  after a time of order $M\ln M$ the
source of thermal radiation shifts gradually  to the star itself. This
allows for a   unitary evolution except possibly for exponentially
small background transition amplitudes. The emerging picture is
compared  with   approaches of t'Hooft and Susskind and the problem of
its overall quantum consistency is evoked.  \vfill\eject

\noindent {\bf 1. Introduction.} \medskip

Black hole evaporation questions acutely the consistency of quantum
physics and general relativity. In the original derivation of black
hole radiance$^{[1]}$, the thermal density matrix describing the
radiation may be viewed as a consequence of tracing pure quantum
states over  states hidden from an external observer by the event
horizon. Disappearance of the black hole would then result in a
violation of unitarity within the universe of external observers, as
suggested by Hawking$^{[2]}$.  A halt of the  evaporation process at
the Planck mass could formally save unitarity by correlating the
distant radiation to a stable planckian remnant$^{[3]}$. The latter
could then be generated in a  huge and probably even infinite number
of ways.  Thus, the end point of the evaporation process would leave
us with the problem either to understand why, despite its breakdown
at the Planck scale, quantum physics is operative at larger scales, or
to enlarge in a consistent way the framework of ordinary quantum field
theory to incorporate infinitely degenerate Hilbert spaces to describe
this Planck scale.

An alternative approach to the dilemma has been proposed by 't
Hooft$^{[4],[5]}$. The Hawking radiation would induce some strong
backreaction on the geometry which would appear, to the external
observer, free of singularity and horizons. The emission process would
be   part and parcel of a  fully unitary evolution of the   black
hole. Susskind and al$^{[6]}$ have suggested to implement this idea by
materializing in the planckian vicinity of the event horizon a
physical ``streched horizon", invisible to the free falling observer,
where the incoming information is deposited and then burned away. The
thermal Hawking radiation would carry mainly coarse grained entropy
and not, as in the conventional picture, entanglement entropy. In this
way   full evaporation would   be consistent with unitarity hidden in
an averaging over   quantum phases. They argue that such a unitary
evolution would imply a departure from hitherto admitted physical
realism as expressed by the concept of an absolute, observer
independent, event.

In absence of further  dynamical ingredients, speculations about the
history of an evaporating black hole rests   largely   on unsafe
preconceptions. The present paper is an attempt to deduce from a
simple and hopefully correct assumption about quantum  gravity a
phenomenological description of the evaporation process.

For all we know, natural phenomenon at all scales presently
encountered  are well described   by physical laws insensitive to
fluctuations above the Planck scale of the classical space-time
geometry in which they operate. It seems therefore reasonable to
assume that, {\it to the extend that a classical background can
operationally be defined}, vacuum fluctuations carrying transplanckian
energy densities cannot have any physical effects. We shall make the
assumption that   if such fluctuations cannot be  gauged away {\it
within a classical background} the dynamics of gravity will alter the
background to reduce them to the Planck scale.   Transplanckian
effects in a background would thus signal gravitational instabilities
and, short of a reliable theory of quantum gravity,  our anzats can
be viewed   as a tentative criterion for selecting   classical
space-times stable against   violent fluctuations of the metric beyond
the Planck scale.

The emergence of   black hole radiance from free field vacuum
fluctuations requires, because of the unbounded blueshift in the
horizons vicinity,
 frequencies well above the Planck scale$^{[1],[7],[8],[9],[10]}$
highly tuned on   distances much smaller than the Planck size. Our
assumption implies that when gravitational nonlinearities are taken
into account, such vacuum fluctuations can be cut-off. This will
reveal background instabilities.

While the very emission of Hawking quanta from free fields does
require sharply localized transplanckian frequencies, these appear
only in the fluctuations of the energy momentum tensor. In addition,
the effect of the Hawking radiation is precisely to render its
expectation value, as felt  by an inertial observer in the vicinity of
the event horizon,  regular and small compared to the Planck
energy$^{[11]}$ . In absence of radiation, this value  would indeed
diverge at the horizon because of the unbounded negative vacuum energy
of matter fields in the  Schwartzschild background. Hawking thermal
radiation thus makes it possible for the free falling observer which
lives on scales large compared to the Planck scale, to experience only
the weak  effect due to the local curvature at the horizon. This
remarkable feature of the Hawking radiation suggests that the
expectation value of the energy momentum tensor, and hence the
average energy density contained in the radiation itself should be
deducible from regularity in the neighbourhood of the horizon without
ever appealing to transplanckian frequencies. We shall see that this
is indeed possible when, according to our assumption, non diagonal
matrix elements are tamed to the Planck scale.

The price to pay for the taming is then not an extinction of the
radiation but rather an instability of classical geometries in the
neighbourhood of the Schwartszchild radius of the star.  For
external observers, the classical geometry  is determined by  an
asymptotically static star imbedded in a hot planckian polarisation
cloud outside the Schwartszchild radius. This is a configuration with
a trivial  topology with neither singularities nor horizons. It
describes a collapse frozen in space and time on the time scale of the
decay.  We shall label such configurations ``achronons" in accordance
with earlier considerations$^{[12]}$. On the other hand, free fall
motion in the planckian vicinity of the Schwartzschild radius opens up
space-time  into the classical black hole geometry and reveal the
masked horizon  and the classical singularity. This picture falls much
into line with ideas  of    t' Hooft$^{[5]}$ and Susskind and
al$^{[6]}$.  Thus while   external  observers probe achronon
geometries, inertial observers cross the horizon and experience
extended black hole geometries. Achronons and black holes are
classically indistinguishable for external observers but are related
by tunneling in the WKB limit of quantized general
relativity$^{[12],[13]}$.

In section 2 the relevant kinematics of the star motion, where for
simplicity the star is idealized by a shell is laid down;   the
structure of vacuum energy momentum fluctuations   due to a collapsing
shell in absence of gravitational nonlinearities is reviewed. These
are violent fluctuations in contradistinction with the smooth
behaviour of expectation values discussed  in section 3.  These facts
are used in section 4 to uncover the nature of a backreaction
consistent with our assumption. The taming of fluctuations leads to a
general picture for collapsing objects where external observers
interpret history as the evaporation of  a heated achronon while free
falling ones cross a horizon and travel towards their destruction in
conventional black hole geometry. The connection with the approach of
references [5] and [6] is made. Unitarity and the possible relevance
of these considerations for quantum physics is touched upon in section
5. \bigskip  \noindent {\bf 2. Fluctuations of the Energy-Momentum
Tensor.} \medskip  Consider a spherically symmetric star of mass $M$
idealized by concentrating its mass at the surface, or equivalently by
a shell of the same mass. Whenever formal developments are needed we
shall use this idealization$^{[14],[15]}$ as it  simplifies the
mathematics without affecting qualitatively the conclusions. Instead
of specifying the energy momentum of the shell and deducing from it
the trajectory, one takes the latter as the input of the analysis.

Let us first characterize the motion of a shell which never crosses
its Schwartzschild radius. Inside the shell, Minkowski space can be
described by the coordinate system $$\eqalign{&ds^2= dU\ dV -r^2
d\Omega^2 \cr &U=\tau - r \qquad V=\tau + r,}\eqno (1)    $$ for $r <
r_s(\tau)$ where $r_s(\tau)$ is the shell radius at time $\tau$. The
input parameter will be taken to be $w_s$ , the radial inwards
velocity of the shell with respect to an observer at rest inside the
star, namely      $$  w_s = -{d r_s(\tau) \over d\tau}.\eqno (2)   $$
Outside the shell, we shall use tortoise coordinates $$ \eqalign {
&ds^2 = \left(1- {2M\over r}\right) du\  dv - r^2 d\Omega^2 \cr &u=
t-r^* \qquad v=t+r^*\cr &dr = (1- {2M\over r}) dr^* ,}\eqno (3)   $$
where $r$, understood as a function of $v-u$, is the ``radius" which
measures the invariant surface $4\pi r^2$ of a sphere and is
continuous across the shell. We have used, inside the shell,   a time
$\tau$ different from the Schwartzschild time $t$,  but one can always
synchronize  the time inside and outside the shell by parametrizing it
everywhere by a single time $t$ continuous across the shell; then
inside,we shall write $$ d\tau = \sqrt{\tilde g_{00}(t)}
dt.\eqno(4)     $$ In general $\tilde g_{00}(t)$ is different from the
lapse function $g_{00}(t) \equiv 1-2M/r_s(t)$ on the outside vicinity
of the star surface and implies a   discontinuity of the red shift
across the star when $w_s \neq 0$. This discontinuity arises because
of the mismatch of inertial frames across the shell and would be
smoothed off as a function of $r$ and $t$ in a regular star. We shall
label  on the star surface $\tilde g_{00}(t)$ and $ g_{00}(t)$
respectively simply by $\tilde g$ and $g$.

We parametrize the whole space-time  by a single $(u,v)$ coordinate
system, performing inside the shell a coordinate transformation $U(u),
V(v)$. Continuity of $ds^2$ and $r$ across the shell together with
Eq.(4)  imply on the shell $$ \eqalignno {{dU\over du}{dV\over
dv}&=g&(5)  \cr dV-dU &=g(dv-du)&(6)
   \cr dV+dU &={\tilde g}^{1\over2} (dv+du) &(7)}               $$
from which one gets $$\eqalignno {{du\over dU} &={1\over
1+w_s}\left({\tilde g}^{-1/2}+w_s g^{-1}\right)&(8) \cr   {dv\over dV}
&={1\over 1-w_s}\left({\tilde g}^{-1/2}-w_s g^{-1}\right)&(9) \cr
g^{-1}&= {1\over 1-w_s^2}\left(\tilde g^{-1} -w_s^2
g^{-2}\right).&(10) } $$  To obtain the $(u,v)$ parametrization inside
the shell, one eliminates  $\tilde g$ in Eq.(10)    and solve Eqs.(8)
and (9) by expressing $w_s $ and $g $  in terms respectively of $u$ or
$v$ only.  In particular for a static shell  we get $\tilde g =
g(r_s)$ where $r_s$ is now the constant radius of the shell; the
redshift is then continuous across the star, and the $(u,v)$
coordinates  inside the shell can be chosen equal to  the $(U,V)$
ones, up to a constant rescaling $\sqrt {g(r_s)}$

We now turn to the collapsing shell and follow the analysis of
reference [10] for the rest of this section. The above description
remains valid  in the restricted space-time available  to an external
observer, that is an observer outside the future horizon at $u=
+\infty$. This is the region depicted in grey in the Penrose diagram
represented in Fig.1. In the limit $u\to +\infty$, $w_s$ tends now to
a constant value $w_s^\infty$ where $0<w_s^\infty\leq 1$. If the shell
collapses along a geodesic trajectory of a Schwartzschild geometry
with mass $M$,  $w_s^\infty = 1/\sqrt 2$ for a   shell  at rest at
$r=+\infty$ and increases towards the limiting value $w_s=1$   with
increasing  initial velocity at $\infty$. On the star surface, we see
from Eqs.(8), (9) and (10)  that $$\lim_{w_s\to w_s^\infty} {\tilde
g\over g^2} =   (w_s^\infty)^{-2}\eqno (11)   $$ and $$\eqalignno  {
&\lim_{w_s\to w_s^\infty}{dU\over du}g^{-1} ={1+w_s^\infty\over
-2w_s^\infty} & (12)\cr
   &\lim_{w_s\to w_s^\infty}{dV\over dv}= {-2w_s^\infty\over
1+w_s^\infty}. &(13)   }   $$ Note that inside the shell   Eq.(12)
still describes the  vicinity of the horizon  whilst Eq.(13) is valid
there only in the limit of the light-like shell $w_s^\infty=1$.
Otherwise  ${dV/ dv}$ is a slowly varying function of $v$ when $r$
goes down from $r_s$ to zero because the line $v=$ constant reaches,
for $r\simeq 0$,  the star surface far away from the horizon in nearly
flat space where ${dV/ dv}$ tends to unity and not to the right hand
side of Eq.(13).   Outside the shell, choosing appropriately the
origin of $r^*$, one may use in the vicinity of the horizon, the
following asymptotic expression for $1-2M/r$   $$1-{2M\over r}
\simeq  \exp\left({v-u \over 4M}\right). \eqno (14)$$ Hence in
Eqs.(11) and (12), we may take $$g =\exp\left({v_\infty-u \over
4M}\right), \eqno(15)$$ where $v_\infty$ is the asymptotic coordinate
of the collapsing shell on the horizon.

Time dependent metrics create particles out of vacuum fluctuations;
for a collapsing shell, once the asymptotic regime  characterized by
$u=O(M)$ in Eq.(15) is reached, particle emission from vacuum tends to
the universal thermal Hawking flux. Let us review how this comes about
and how transplanckian frequencies enter the game, taking for
simplicity  a light like trajectory\foot{This does not affect any of
our order of magnitude estimates.}:  Eqs.(12) and (13) can then be
integrated to yield inside the shell $$\eqalignno{
U&=-4M\exp{v_\infty-u\over 4M} &(16)\cr V&=v. &(17)}$$

We first consider the s-wave contribution from a free scalar field to
the radiation.  Moving  back in time, positive  frequency plane waves
in the retarded time $u$ defined on ${\cal I}^+$   reflect on the
timelike  curve $r=0$ and propagate to ${\cal I}^-$. The curve  $r=0$
can also be labeled by $U=v$ and hence the s-waves arriving on ${\cal
I}^-$ span only   the negative $v$-axis. Hence wave-packets build out
of positive frequencies in the retarded time on ${\cal I}^+$ require
frequencies of both signs in the advanced time $v$ on ${\cal I}^-$. To
build a complete set of solutions of the field equations with positive
frequencies on ${\cal I}^-$, one needs waves having support on the
positive $v$-axis and propagating towards the horizon. The Heisenberg
vacuum  $\ket{0}$ annihilated by destruction operators associated with
positive frequencies on  ${\cal I}^-$ is then a superposition of pairs
formed from outgoing quanta on ${\cal I}^+$ correlated to partner
states describing waves propagating towards the horizon $\cal H$. Note
that while the outgoing quanta describe real particles, this is not
true for their partners which have no definite frequency sign and
should be interpreted as vacuum fluctuations. Taking a trace of  $\ket
{0}\bra{0}$ over the latter states, one gets a density matrix
describing a thermal flux of outgoing quanta on ${\cal I}^+$ at the
temperature $$ T = {1\over 8\pi M}. \eqno (18) $$

We depict a typical quantum or ``Hawking photon"  arriving  at the
retarded time $u_0$ on ${\cal I}^+$ by a   wave-packet extending over
a distance of the order of a wavelenght $M$. Its complex frequency
spectrum $f(\omega)$ has the form $$f(\omega) =\vert f(\omega)\vert
\exp (-i\omega u_0) \eqno(19)$$ where $\vert f(\omega)\vert$ is
centred around a positive frequency $\omega_0$  of order $T$ and
spread over a comparable range. We normalize the wave packet by
$$\int_0^\infty d\omega \vert f(\omega)\vert^2 =1\eqno(20)$$
 When the wave packet moves back in time with the velocity of light,
it   experiences upon penetrating inside the shell a differential
blueshift $ dU/ du = -U/ 4M$ with respect to a   Minkowskian observer
at rest inside the shell. The blueshifted central frequency
becomes      $$\tilde\omega_0 =\omega_0\tilde g^{(-1/ 2)}= \omega_0
g^{-1} = 4M \omega_0 / \vert U \vert. \eqno(21) $$ The packet then
propagates    with the same frequency to ${\cal I}^-$ after reflexion
on $r=0$. For definiteness  we take $\omega_0 = 1/2M$. In the
asymptotic regime this is $$ \tilde\omega_0 = \eta^{-1} \qquad
\eta\equiv r_s(\tau) -2M \eqno (22)$$ where $r_s(\tau)$ is the radius
at which a light ray $u=$constant intersects the shell. When $\eta$
becomes smaller than a Planck length, $\tilde\omega_0$ crosses the
Planck scale. Eq.(14) shows that one needs only a retarded time $$ u_0
\simeq M\ln M \eqno (23)$$ to reach this Planck scale. For times of
order  $M^3$, thus comparable to the black hole lifetime, one would
get $\eta \simeq M \exp (- M^2)$, and $\tilde\omega_0 \simeq \
M^{-1}\exp M^2$. This increase in frequency   is accompanied by   a
localisation within a wavelenght $\tilde\omega_0^{-1}$. The
wave-packet is correlated on ${\cal I}^-$to a partner centred at a
positive $v$ value. This correlation arises because a plane wave of
frequency $\tilde\omega_0$  extends on both positive and negative $v$
values. The partner with $v>0$ can then be similarly depicted as a
wave packet with the same frequency as the ancestor of the Hawking
photon but the separation between the two packets is on  ${\cal I}^-$
comparable to their spread.

We thus arrive to the following qualitative description of the
generation of an s-wave  Hawking photon out of free fields (see
Fig.1).   Select at $t=-\infty$ on $\cal {I}^-$   among the vacuum
fluctuations of empty Minkowski space, two spherically symmetric
distributions  moving with the velocity of light. The outer
fluctuation (the partner)  never reaches $r=0$ for the external
observer. After passing the curve $r=0$ the inner fluctuation (the
Hawking photon ancestor) meets the outer one and then separates,
propagates towards  $\cal {I}^+$ and converts to a real quantum by
redshifting its   frequency  $\tilde\omega_0$.  The real quantum is
still correlated to the fluctuation moving towards the event horizon.
The local frequency $\tilde \omega_0$ of the  pair is determined by
the redshift required to realize the conversion of the ancestor into a
real quantum of frequency $\omega_0$, or equivalently by energy
conservation. The distance between the two poles before their
separation is roughly $\tilde \omega^{-1}$ and   of the order of their
spread. For quanta emitted a very short time  after the start of  the
Hawking emission, namely after a retarded time $u_0$ given by
Eq.(23),  the fluctuations  have very large transplanckian frequencies
and are localized over   distances well below the Planck size in the
frame of the Minkowskian observer at rest inside the shell.

This history of the generation of Hawking photons from vacuum
fluctuations can be expressed in quantitative terms by analysing their
energy content.  One post-select the normalized state vector $\ket
{P_1}$ representing the Hawking photon arriving  on  $\cal {I}^+$ at
the retarded time $u_0$ and computes the state of the correlated
partner $\ket {P_2}$ which is, up to a normalization constant, equal
to $\langle 0 \ket {P_1}$.   The energy momentum of the pair at any
time is then defined interms of the ``weak value"  of the energy
momentum tensor operator  $\hat T_{\mu \nu}(x)$, that is its matrix
element,  suitably normalized,   between the   state $\ket 0$ and the
correlated pair:
 $$ T_{\mu\nu}^{weak}(x)\equiv{\bra{P_1}\bra{P_2}\hat T_{\mu
\nu}(x)\ket 0 \over
 \bra{P_1} \bra{P_2}\ket 0}. \eqno (24)            $$ As shown by
Aharonov and al$^{[16]}$, if a future measurement were to yield the
post selected state,   the real part of the weak value of such a
hermitian operator     can be recorded  by a ``weak" non demolition
measurement and  its imaginary part induces a shift of the conjugate
variable of the measuring device. A general analysis of the energy
content of vacuum fluctuations in terms of weak values was given in
reference [9].

One first computes $T_{uu}^{weak}(x)$ on  $\cal {I}^+$ and one
verifies  that  the total weak energy   contained in the wave-packet
is  equal to the total average energy  $<\omega>$ of the emitted
quantum$^{[10]}$. One gets indeed $$\int du \lim_{v \to +\infty}
T_{uu}^{weak}(x) = <\omega>= { \int d\omega  \exp (- \omega 8\pi M)
\vert f(\omega)\vert^2 \omega\over
 \int d\omega
    \exp (- \omega 8\pi M)
   \vert f(\omega)\vert^2 }.    \eqno(25)$$ Here $f(\omega)$ is the
spectral function defined in Eqs.(19) and (20).   Note that the
average is taken not only with respect to the quantum
 weight $f(\omega)$ but also over a thermal distribution at the
Hawking temperature $(1/ 8\pi M)$.  The distribution is here
bolzmannian because we have  post selected a single pair. Tracking
back in time $ T_{uu}^{weak}(x)$ along   light-like curves $u
=$constant we penetrate   inside the shell. There, in the empty
Minkowski space-time, the wave packet energy in terms of
$T_{UU}^{weak}(x)$ is related to $ T_{uu}^{weak}(x)$ by the
differential blueshift $ dU/ du = -U/ 4M$ and is conserved after
reflexion along light-like curves $v=$constant up to $\cal {I}^-$.
Explicit evaluation of $ T_{vv}^{weak}(x)$ on $\cal {I}^-$ recovers
for negative $v$ the blueshifted wave packet described by
$T_{UU}^{weak}(x)$ and exhibits, for positive $v$ the wave packet
energy of the partner. The total weak energy on $\cal {I}^-$ is
 $$\int dv \lim_{u \to -\infty} T_{vv}^{weak}(x) = 0.\eqno(26)$$ This
result is in accordance with the fact that $\ket 0$ is an eigenstate
of zero total Minkowski energy and thus that the weak energy
characterizes on  $\cal {I}^-$ vacuum fluctuations. This compensation
is made possible because, while the real weak energy of the partner is
positive , the real weak energy of the   ancestor has become negative
due to the differential blueshift. Indeed   the post selection of a
packet of finite extension induces small energy oscillation tails and
the huge blueshift of a negative energy tail close to $U=0$
overcompensates the large but smaller blueshift of the positive core
energy of the Hawking photon. Both the partner energy and the core
energy of the ancestor are correctly given by Eq.(22).

Thus the qualitative picture of the generation of Hawking photons is
entirely confirmed and the pair of fluctuations selected on $\cal
{I}^-$ to describe the ancestor and its partner form a dipole in
energy. The outer pole (the partner) carries positive Minkowskian
energy while the inner pole ( the ancestor) has a total negative
energy which will later turn into a real quantum by losing its tail
and redshifting its positive core energy. These dipoles carry thus
huge transplanckian energy carefully tuned on  correspondingly small
distances {\it inside} empty Minkowsky space.

It is important to realize that it is impossible to reduce these
transplanckian dipoles to planckian ones by some gauge transformation
in the given background. To do so, one would have    to scale
$\tilde\omega_0$ down to the Planck scale everywhere on the dipole
spheres and between them so that their separation $
\tilde\omega_0^{-1}$ would be stretched accordingly to a Planck
length. This cannot be achieved through local Lorentz transformations
without reintroducing transplanckian Unruh effects. Global Lorentz
transformations are not available either. They would amount to
multiply the right hand side of Eq.(8) by a constant $\lambda \neq 1$
and of Eq.(9) by $\lambda^{-1}$. This is not possible because the
continuity of $r$ at the star boundary  fixes $\lambda$ to be unity.
The invariant meaning of $\tilde\omega_0$ is explicitly related to the
existence of the collapsing star. This should be contrasted with
the dependence of minkowskian frequencies on   Lorentz boosts in the
description of the Hawking radiation from the Unruh vacuum where  the
effect of the star collapse is mimicked by suitable boundary
conditions on a fictitious past horizon. The latter description
introduces a spurious symmetry  which is just such a Lorentz boost
and which is equivalent
 to a global Killing symmetry $t \to t+$constant. This global symmetry
is clearly broken inside the
 collapsing shell.

Finally,  we point out that the above restriction to s-waves and to
non interacting fields do not change these conclusions as long as
gravitational non linearities are not taken into account. For free
fields, higher angular momenta waves inside the shell still propagate
at the high frequency $\tilde\omega_0 = \eta^{-1}$   and are therefore
quite insensitive to the centrifugal barrier $l(l+1) / r^2$ there
until they reach  very small distances from the centre of the star (
$r \leq \sqrt{2\eta M}$) whereupon they are reflected. The above
analysis remains essentially valid for all angular momenta except that
most of the high angular momentum modes are reflected back outside the
star  towards the horizon. Transplanckian dipoles exist for all
angular momenta but   outside the shell the distribution of Hawking
photons ancestors is the result of a balance of outgoing and reflected
waves.  This brings them  nearly in thermal equilibrium close to the
horizon outside the star  with a local blueshifted temperature $$
T_{loc} = {1\over 8\pi M}(1-{2M\over r})^{-1/2}
 \eqno(27)$$ of the order of the blueshifted frequency there. Thus,
the inclusion of higher angular momentum for free fields does not
change the transplanckian character of the production process. Rather,
it imbeds, outside the shell, the s-wave transplanckian frequencies in
a transplanckian thermal bath, which at this stage is essentially
kinematical as it does not rely on interactions. Interactions due do
asymptotically free interacting renormalizable field theory mixing
different angular momenta of single hot photons will be weak at the
Planck energy and could only contribute to stabilize the temperature.
Hence  they cannot
 reduce the frequencies $\tilde\omega$ to planckian   values.

The above detailed description of the Hawking photon emission rests
entirely on the non diagonal elements of the energy momentum tensor
which contribute to correlation functions such as $\bra 0\hat T_{\mu
\nu}(x)\hat T^{\mu\nu}(y)\ket 0$. These non diagonal elements   are
related to the diagonal ones  $$    T_{\mu \nu}(x)  \equiv \bra 0\hat
T_{\mu \nu}(x)\ket 0\eqno (28)$$ by $$    T_{\mu \nu}(x) =
\Sigma_{P_1,P_2} \langle 0\ket{P_1}\ket{P_2}
\bra{P_1}\bra{P_2}\hat T_{\mu \nu}(x)\ket 0 = \Sigma_{P_1,P_2}\vert
\langle 0\ket{P_1}\ket{P_2}\vert^2 T_{\mu\nu}^{weak}(x)\eqno(29)$$
where the sum is over a complete   set formed from the tensor product
of states  $\ket{P_1}$   defined   on $\cal {I}^+$ with their
correlated partners $\ket{P_2}= \bra{P_1}0\rangle$.  In empty space,
here $v<v_s$, $     T_{\mu \nu}(x) $ is zero in sharp
contradistinction to the   weak energy of the fluctuations generated
from the post selected state.
  Eq.(29) means   that the expectation value of the energy-momentum
tensor, which is zero on $\cal {I}^-$, vanishes there because of
destructive interference between elementary emission processes.
\bigskip \noindent {\bf 3. The Average Energy-Momentum Tensor.}
\medskip Before coping with the transplanckian fluctuations of the
field vacuum in the metric of a collapsing star, it is important to
understand the behaviour of the  expectation values of the
energy-momentum tensor for an arbitrary motion of the shell.   Four
dimensional computations for free field in black hole geometries with
suitable boundary conditions indicate that all the components of   the
energy-momentum tensor have finite and small (on the order of the
local curvature) expectation values on the future horizon in the
inertial frame of a free falling observer$^{[11]}$. This is a
consequence of the {\it average} Hawking flux in the horizons
vicinity. In its absence, as would be the case  if one would impose
the vanishing field boundary conditions there, divergences do appear.
The latter ``Boulware" boundary condition  is the natural one for a
classical star at rest infinitesimally close to its Schwartzschild
radius. These results are not qualitatively affected if one restrict
the field to depend only on the $(r,t)$ two dimensional subspace of
the full metric Eq.(3) and can then be derived analytically from the
trace anomaly$^{[17]}$. We shall follow this method here for a general
motion of the shell and extend it to take into account a possible
interaction between the field and the star. This will  be an essential
element of our approach to the back reaction problem.

We write the metric due to the spherical shell in the two dimensional
subspace of the   $(u,v)$ coordinate system as  $$  ds^2 = e^{2\rho}
du  dv. \eqno(30)$$ Outside the shell, the metric factor $e^{2\rho}$
is independent of the shell motion and is equal to $  1- (2M/ r)$, but
its value inside is contingent upon this motion. For   a generic
collapsing
 shell, in the vicinity of the horizon and of the shell, one has from
Eq.(15) $$ \rho = {v_\infty-u \over 8M} \eqno(31) $$ while for a
static shell of radius $r_s$, $\rho$ takes the constant value $$\rho=
1- {2M\over r_s} .\eqno(32) $$ Conservation of the (2-dimensional)
energy momentum tensor $\hat T_{\mu \nu}^{(2)}(x)$   $$ \hat
T_{\nu;\mu}^{\mu\,(2)}(x)  =0\eqno(33)$$ can be written, using in
Eq.(33) the trace anomaly of a free scalar field $$ T_{uv}^{(2)} =
-{1\over 12\pi}  \partial_u \partial_v \rho \eqno (34) $$ $$\eqalign {
\hat T_{uu}^{(2)}&= {1\over 12\pi} \left[ \partial^2_{uu} \rho -
(\partial_u\rho)^2 \right] + \hat t_u(u) \cr \hat T_{vv}^{(2)}&=
{1\over 12\pi} \left[ \partial^2_{vv} \rho - (\partial_v\rho)^2\right]
+ \hat t_v(v) .}\eqno(35)$$
 It is important to realize that while the trace anomaly in Eq.(34) is
a c-number, the integration functions $\hat t_u(u)$ and $\hat t_v(v)$
in Eqs.(35) are genuine explicitly conserved operators. It is their
non diagonal conserved matrix elements which gave rise to the
transplanckian generation of radiation quanta. Here we shall focus our
attention to the  expectation values $T_{uu}$ and $T_{vv}$ which are
combinations of non tensor expectation values $t_u$ and $t_v$ with the
non tensor c-numbers in Eq.(35).  Note  that in the  2-dimensional
approximation of the correct 4-dimensional  $\hat T_{\mu \nu}(x)$ one
should use the correspondence  $ \hat T_{\mu \nu}^{(2)}(x) = 4\pi r^2
\hat T_{\mu \nu} (x)  $.

Outside the star, we get from Eq.(35) $$\eqalignno {  T_{uu}^{(2)}&=
{1\over 12\pi} \left[ -{M\over 2r^3}(1-{2M\over r}) - {M^2\over 4r^4}
\right] +   t_u(u) &(36)\cr  T_{vv}^{(2)}&= {1\over 12\pi}  \left[
-{M\over 2r^3}(1-{2M\over r}) - {M^2\over 4r^4} \right] +   t_v(v)
&(37)\cr   T_{uv}^{(2)}&= -{1\over 12\pi} {M\over 2r^3}(1-{2M\over r}).
 &(38)}$$ The functions $t_u(u)$ and $t_v(v)$ depend on the boundary
conditions of the fields.  Vanishing of the energy-momentum tensor on
$\cal {I}^-$ implies from Eq.(37) $$t_v(v) = 0 .\eqno(39)$$
 We shall see that $t_u(u)$ is then entirely determined from the
motion of the shell.  Inside the shell Eq.(35) expressed in the
Minkowskian coordinates Eq.(1)  gives  $$ T_{UU}^{(2)} = t_U(U)
\qquad  T_{VV}^{(2)} = t_V(V) \eqno(40)$$ and the boundary condition
at $r=0$ imposes $$ T_{UU}^{(2)} + T_{VV}^{(2)} = 0
\quad\hbox{at}\quad U+V=0. \eqno (41) $$ It follows from Eq.(39), (40)
and (41) that the transformation properties of the functions $t_u(u)$
and $t_v(v)$ under coordinate transformations, as applied between the
$(u,v)$ and the $(U,V)$  coordinates inside the shell, entirely
determine $t_u(u)$.

Let in general   $(\tilde u, \tilde v)$ be some conformal
reparametrisation defined in some neighbourhood. The conformal factor
$e^{2\rho}$ in Eq. (30) transforms according to  $$ \tilde\rho = \rho
+ {1\over 2}\ln  u'+ {1\over 2}\ln  v'\eqno(42)$$ where $u'  \equiv
du/ d\tilde u; v'  \equiv dv/ d\tilde v   $. From the tensor
transformation properties of $ T_{uu}^{(2)}$ and $ T_{vv}^{(2)}$ under
conformal reparametrization and from Eq.(42) one gets  $$ t_u(u) =
{1\over 12\pi} \left[ {u'''\over 2 u'^3} - { 3u''^2\over 4 u'^4}
\right] + {1\over u'^2} t_{\tilde u} (\tilde u) \eqno(43) $$ and
similarly for $t_v(v)$ in terms of $v'$. If  $(\tilde u, \tilde v)$ is
taken to be the $(U,V)$ system inside the star, $u'$ and $v'$ are
computable in terms of the shell motion from Eqs.(8),(9) and (10).

For the light-like collapsing shell in the asymptotic region Eqs.(17)
and (43) give $t_V(V)=0$ and hence from Eq.(41) $t_U(U)=0$. Using
again Eq.(43) and the coordinate transformation Eq.(16), one gets for
$t_u(u)$ the constant value $$   t_u = {\pi\over 12} {1\over
(8\pi M)^2}.\eqno (44) $$ It is easy to verify, using Eqs. (11), (12)
and (13), that in the asymptotic regime, the result Eq.(44) does not
depend on the light-like limit and is true for generic collapsing
shell.

This result implies that the energy flux across any sphere
$r=$constant outside the shell has the constant value  $$
T_{uu}^{(2)}- T_{vv}^{(2)}= t_u \eqno(45)$$  as given by Eq.(44). This
is, as expected, the thermal Hawking s-wave flux at the Hawking
temperature Eq.(18). On the other hand, for a static shell, no
radiation can occur from the free field vacuum and Eqs.(43) now yields
$t_u=0$, confirming the absence of energy flux outside the star. On
the horizon, outside a collapsing shell, the energy flux
$T_{vv}^{(2)}$ crossing the horizon is, from Eqs.(37) and (39),
finite, negative and equal in absolute value to the Hawking flux
$t_u$. In contradistinction,  $T_{uu}^{(2)}$ vanishes on the horizon
as a consequence of Eqs (36) and (45). The negative energy flux is
thus necessary to ensure energy conservation in presence of the
Hawking radiation and is sometimes referred to as the source of this
radiation. Note however that, for a static star close to its
Schwartzschild radius $T_{vv}^{(2)}$ tends in its neighbourhood to the
same negative value. The difference in the static
 case  is that this negative flux is compensated by an equal amount of
negative  $T_{uu}^{(2)}$. It is the latter quantity and not
$T_{vv}^{(2)}$ which is affected by the motion of the shell, being
brought to zero on the horizon by the positive constant contribution
$t_u$. It is easy to verify  that   $T_{uu}^{(2)}$ tends to zero as
$(1-2M/r)^2$. This is what is needed to ensure that all the components
of the energy-momentum tensor are finite in the inertial frame of the
free falling observer. The Hawking energy has thus the important
property of ensuring that the free falling observer does not
experience singular contributions of the field vacuum as he approaches
the horizon of a collapsing star, a feature which would not be true if
the classical star would be ultimately brought to rest.

The whole preceding analysis is   contingent upon the validity of the
conservation law Eq.(33). There, it was implicitly assumed that the
star and the scalar field are not interacting and therefore obey
separate conservation laws in the background space-time. However, as
will be discussed in section 4 and 5, this is a non trivial assumption
when the shell is  within a Planck size of its Schwartszchild radius
when gravitational backreaction is taken into account. To pave the way
for studying this problem,  we shall enlarge   Eq.(33) to include a
possible source term of  the field energy-momentum tensor located at
the classical position of the shell. We limit our considerations to
expectation values.

We thus generalize Eq.(33) to  $$  T_{\nu;\mu}^{\mu\,(2)}(x)  =
J_{\nu}(x) \eqno(46)$$ where $J_{\nu}(x)$ represents the source term.
The previous analysis basedon the trace anomaly equation is still
locally valid but the functions $t_u(u)$ and $t_v(v)$ need no more be
the same inside and outside the shell. To make this explicit we
express the source current  in the global $(u,v)$ coordinate system as
 $$\eqalignno{ J_u(u,v) &= 2 e^{-2\rho} k_u(u) \delta [v-v_s(u)] &
(47)\cr J_v(u,v) &= 2 e^{-2\rho} k_v(v) \delta [u-u_s(v)] & (48)}$$
where $v_s(u)$ and $u_s(v)$ are parametrizations of the shell
trajectory. The functions $k_u(u)$ and $k_v(v)$  will measure the
radiation of the shell.

To see this, define a function $F(u,v)$ which goes through zero on the
shell and is positive outside. We have $$ \partial_v\Theta [F(u,v)] =
\delta [v-v_s(u)];\quad \partial_u\Theta [F(u,v)] =-\delta [u-u_s(v)].
\eqno(49)$$ Using Eq.(49), the solution of Eq.(46)   with the source
terms Eqs.(47) and (48) is now, using the trace anomaly Eq.(34),
$$\eqalign {   T_{uu}^{(2)}&= {1\over 12\pi} \left[ \partial^2_{uu}
\rho - (\partial_u\rho)^2 \right] +  t_u(u) +   \Theta [F(u,v)] k_u(u)
\cr   T_{vv}^{(2)}&= {1\over 12\pi} \left[ \partial^2_{vv} \rho -
(\partial_v\rho)^2\right] +   t_v(v)  - \Theta [F(u,v)]
k_v(v)}\eqno(50) $$ The vanishing of the flux on  $\cal {I}^-$  now
implies $$ t_v(v) - k_v(v) =0 \eqno(51)$$  and the energy flux across
a sphere of radius $r$ outside the sphere at the retarded time $u$
becomes in general $$ T_{uu}^{(2)}- T_{vv}^{(2)}= t_u(u)+k_u(u).
\eqno(52)$$

As previously, $t_u(u)$ and $t_v(v)$ transform under conformal
reparametrization according to Eq.(43).  The functions $k_u(u)$ and
$k_v(v)$ which give the possible genuine contribution of the shell
to   radiation transform  as $$ k_{\tilde u}(\tilde u) = u'^2
k_u(u);\quad k_{\tilde v}(\tilde v) = v'^2 k_v(v). \eqno(53)$$ One
verifies indeed that Eq.(53) ensures the correct transformation
properties of the source currents and of the components of the
energy-momentum tensor.     \bigskip \noindent {\bf 4. Backreaction
and Radiance in the Tamed  Vacuum.}  \medskip The thermal   radiation
in the background of a collapsing shell is characterized by an energy
momentum tensor whose expectation value is smooth  but whose
fluctuations are exponentially large compared to the Planck scale and
are correlated over distances exponentially smaller than the Planck
size. Classical general relativity is not likely to make any sense at
those energies and space-time separations and therefore     the back
reaction due  gravitational interactions seem to depend on at least
some genuine properties of quantum gravity. Nevertheless the smooth
character of expectation values has nourished the hope that a
semi-classical approach to backreaction whereby these average
$T_{\mu\nu}$ are taken as a source of the classical Einstein
equations,  is valid for macroscopic black holes.

This conventional point of view means that as long as the evaporating
black hole does not reach the Planck size, gravitational backreaction
is weak. Except at the final stage, one could treat in good
approximation the emission process adiabatically by fixing the
classical Schwartzschild background by the instantaneous mass. The
cumulative effect of the varying mass leads however to  qualitative
effects$^{[18],[19]}$ which we now discuss.

For a constant mass the global event horizon coincided  outside the
star with the boundary of trapped surfaces, that is with the apparent
horizon. The latter now separates from the former and spans a
time-like surface depicted in Fig.2. The proper time necessary for the
shell to move from the apparent horizon to the event horizon is of the
order of the Planck time but this motion appears to be of the order of
the black hole lifetime for the distant observer. A hypothetical light
ray which would be emitted from inside the star and cross the shell
between the two horizons would first recede to smaller radius and
diverge again upon reaching the apparent horizon. This means that the
evaporation of the black hole is not due to a reduction of the shell
mass but results from a negative energy cloud which originates in the
negative flux $T_{vv}$ compensating in the static case the energy
carried away by the Hawking radiation. This cloud is situated at
smaller  radius than   the shell when the latter was in causal contact
with the distant observer and is separated by a space-like distance
from the shell at the same radius. Thus the march of the classical
shell  towards its final destruction  by tidal forces in the vicinity
of the classical singularity appears as a causally disconnected
history from the evaporation out of a polarisation cloud which fills
up with increasing mass (in absolute value)  a macroscopic volume of
space ``outside the star surface" back to $r\simeq 0$.

This bizarre sequence of events is fully consistent with  a thermal,
structureless Hawking radiation encoding no information about the
original star. All the information is contained in the collapsing star
evolving with its initial mass. This is also consistent with the fact
that, as required by the quantum superposition principle, there is no
duplication of information between the collapsing star and the Hawking
radiation$^{[6]}$. In fact, the latter argument seem to indicate that
the picture emerging from the semi-classical backreaction anzats is
bound to survive quantum corrections and that up to the Planck mass
scale, the Hawking radiation of a collapsing star cannot, even in
principle, contain any relevant information about the detailed
structure of the star. One would then be confronted with the usual
dilemma of either a full evaporation and the concomitant loss of
unitarity  in the universe of external observers or with the halting
of the collapse at the external observers Planck size, relegating
there the huge and presumably infinite degeneracy of a left over
remnant.

However, we have to examine whether the above picture is reasonably
consistent with the existence of the large fluctuations of the energy
momentum tensor. Clearly the transplanckian fluctuation spectrum does
not satisfy the background stability criterion introduced in section
1. Cutting off the dipolar fluctuations at the Planck size on $\cal
{I}^-$ would suppress the mechanism reviewed in section 2 responsible
for the generation of Hawking radiation as seen by the external
observer. According to our   assumption this should imply that either
this mechanism for black hole radiance is qualitatively altered when
gravitational non linearities are taken into account or that   Hawking
radiation does not occur.

There are at least two reasons why the  latter alternative seems
unreasonable. First the  thermal   radiation appears as a general
thermodynamic property of the hole if one admits the Bekenstein
conjecture$^{[20]}$ that  the area of the event horizon is a measure
of entropy.  To see this, recall that the black hole entropy must then
be, for dimensional reasons, inversely proportional to the Planck
constant. This in turn requires that an eternal black hole of mass $M$
surrounded by static matter with total mass at infinity $M_\infty$
should have a global temperature proportional to $\hbar$. Consider
indeed the classical Killing identity$^{[21]}$ (it can be viewed as
the integrated constraint equation over a static coordinate patch)
which can be written as   $$-{\kappa \over 2\pi} \delta {A\over 4}=
\delta H -\delta M_\infty \eqno(54) $$ where $\kappa$ is the surface
gravity of the hole  and $\delta H$ is the variation of all non
gravitational parameters in the matter hamiltonian outside the
horizon. Eq. (54 )   permits   the identification of a multiple of the
surface gravity to a temperature. The equation is classical and thus
the temperature should be proportional to $\hbar$ to cancel the
$\hbar^{-1}$ in the entropy. This argument gives credence to the
derivation and even to the estimate of the temperature via euclidean
continuation of the metric, either for Green's functions, for
partition functions$^{[22]}$  or for tunneling amplitudes$^{[13]}$,
because euclidean continuation always leads to the required dependence
on $\hbar$ and because all these methods yield the same result, namely
the Hawking temperature Eq.(18)for black holes with no surrounding
matter. The fact that this thermodynamical argument refers more
directly to hypothetical eternal black holes than to incipient ones
does not weaken its significance. The  production of Hawking quanta
from   vacuum fluctuations can be viewed as arising from the local
thermalization  close to the horizon at the temperature Eq.(27) and
this temperature is the same as in the eternal black hole case.

Second, as was stressed in section 3,  Hawking radiations contribution
to the average energy density is needed  to render finite and small,
in a free field theory, the vacuum expectation value of the
energy-momentum tensor experienced by a free falling observer
crossing the horizon . Cutting off transplanckian effects in the
computation of expectation values would only confirm this result when
gravitational effects are taken into account. Smooth expectation
values are certainly required for retrieving the classical motion of
the free falling observer  confronted only with the small local effect
of the curvature at the horizon; they provides therefore a strong
argument in favour of the Hawking radiation.

We are thus led to conclude that  Hawking radiation should exists and
therefore that gravitational non linearities set in and qualitatively
modify the emission process  after the characteristic  retarded time $
u_0 \simeq M\ln M$ defined by Eq.(23) where transplanckian effects
appear inside the shell. At that  time the radial distance separating
the shell from the horizon is of order unity according to Eq.(22) but
the local temperature outside the star is  $$ T_{loc}=
O(M^{(-1/2)})\eqno(55)$$ as seen from Eq.(27) and hence far below the
Planck temperature for a macroscopic black hole. The reason for the
mismatch between frequencies outside and inside the shell is the
additional blueshift induced by its motion as encoded in Eq.(11). This
is however not an artifact of the shell model because in a realistic
star it would simply be spread on a distance of the order of the star
radius.

To understand the nature of gravitational effects, we thus cut off
transplanckian effects for times  greater than $u_0$. This
determines,  according to our    assumption, the stable  geometry in
which the Hawking radiation must take place. The resulting  geometry
is topologically equivalent to Minkowski space-time because   the
shell slows down and stops when the local temperature outside the
shell   reaches  the Planck temperature. This will now be exemplified
in the 2-dimensional approximation for a light-like shell collapse but
the qualitative conclusions are valid in general.

When $\tilde\omega_0$ reaches unity, a Hawking quantum of say $\omega_0
=1/2M$ in (21) has been blueshifted by the factor   $\tilde g^{-1/2}$
which reaches     $ 2M $. If we fix for $u\geq u_0$ $\tilde g$ to the
corresponding value in Eq.(10), we get  in the asymptotic region a
differential equation for the motion of the shell $$ \left(
{d\eta\over d\tau}\right)^2 = \eta \left(\eta-{1\over 2M}\right)\qquad
\eta \geq {1\over 2M}. \eqno (56)$$ Thus, at $\eta =1/2M$, the shell
is brought to a halt. The outside temperature has reached the Planck
scale, as seen from Eq.(27) and the proper distance of the shell to
the Schwartszchild radius at fixed the external time $t$, is then also
of order unity. The slowdown time $\Delta\tau$,
  is from Eq.(55) $$\sqrt{2M} = \cosh {\Delta\tau\over2}. \eqno (57)$$
and takes thus only a few Planck times in the $\tau-$time.

During this time span  the emission process is gradually altered and
after the time $\Delta\tau$ the original mechanism is completely lost
since the star has come to rest. The only way radiation  can occur is
from the Planckian polarisation cloud in which the shell is immersed
and this requires quantum transitions between the constituents of the
shell and its Planckian cloud. According to the above discussion we
have to assume that  in an   average sense, the radiation is still
thermal at  the global temperature Eq.(18).

Phenomenologically, this can be accounted for in the 2-dimensional
s-wave approximation by the source term $k_u(u)$ Eq.(47) constituted
by our polarized star. The full energy flux is $t_u(u)+k_u(u)$ as
given by Eq.(52); for $u<u_0$ it is all contained in $t_u(u)$ and
gradually shifts to the source  term $k_u(u)$ as the shell slows down
and stops. The total flux sticks to the Hawking value Eq.(44) when
$t_u(u) \to 0$ and we get $$t_u(u)+k_u(u)={\pi\over 12} {1\over
(8\pi M)^2}\eqno(58)$$ which determines $k_u(u)$ as a function to the
radial distance to the horizon.  Note that regularity alone in the
vicinity of an eventual horizon in the inertial frame  fixes
uniquely the right hand side of Eq.(58) independently of any
thermodynamical assumption: a $u-$dependent term $t_u(u)+k_u(u)$ can
annihilate the $r-$dependent term in Eq.(36) only if it is constant
and therefore equal to the right hand side of Eq.(44).

The picture that emerges from our   assumption when gravity enters the
game is that Hawking radiation occurs within a background which is
free of horizon and singularity. Its source shifts gradually from the
vacuum fluctuations in the metric of a collapsing shell to that of a
static shell imbedded in a hot planckian polarisation cloud. This
configuration of the shell or more generally of any star is not
distinguishable for an external classical observer from a collapsing
object. The object is however  frozen in space-time which means that
the geometry has closed on a trivial topology. Such frozen structures
where the star sticks within a Planck size of its Schwartzschild
radius have been introduced previously and labeled as ``achronons";
they are related to black hole geometries by tunneling amplitudes
$\exp (-A/8)$  in the WKB limit of quantized general relativity. If
the achronon geometry provides the   framework in which external
  observers   perceive the Hawking radiation, it is not fit to the
world of free falling ones in the vicinity of the Schwartzschild
radius which are blind to the radiation. Quantization in their local
frames confirm that they can only experience rare particle creations
due to the small local deviations from global Minkowski space-time.
This should be a result surviving gravitational non linearities
because it does not require any planckian effects$^{[23]}$. The sole
effect of Hawking radiation is precisely, as already emphasized in
section 3, to erase any precursor of the star which could affect free
falling motion by wiping off in its vicinity divergences in
expectation values. Therefore, as a consequence of the Hawking
radiation itself, there appear to be no reason preventing free falling
observer to travel according to the prediction of classical physics.
For them, space opens up in the vicinity of the Schwartszchild radius
of the star as he crosses the masked horizon to reveal the extended
and classically singular black hole geometry. The taming of the vacuum
according to our assumption induces background instabilities which
makes the space-time geometries underlying the history of a collapsing
star dependent on the state of motion of the observer: for external
observers the collapse does not generate a black hole but a heated
achronon while free falling ones do experience a   black hole geometry.

This conclusion is in line with the approach of t'Hooft and Susskind.
However, the scenario developed in reference [5] where the star emits
a large amount of thermal quanta before turning around contradicts our
basic assumption. The model of [5] will unavoidably lead to
uncontrollable transplanckian effects and seems therefore to suffer
from the same potential inconsistency as the original approach. Our
achronon has more similarity with the streched horizon of reference
[6] but it is not only a layer where information is deposited, it is
the star itself interacting with a cloud of  planckian constituents.

\bigskip \noindent {\bf 5. Gravity and Quantum physics} \medskip The
shell model  illustrates clearly the   key difference between the
present approach and the conventional one exemplified by the
semi-classical construction of Fig.2. The shell and its polarisation
cloud responsible for radiative emission have been merged on a  Planck
scale and can thus be depicted in classical terms by a single
trajectory. This would at first sight seem to be harmless and is in
fact consistent with the conventional picture {\it as long as the
black hole does not loose mass}.   Processes  induced by gravity
between the scalar field and the quantum degrees of freedom of the
shell are expected to play an important role in the description of the
Hawking emission once the hitherto neglected backreaction is taken
into account. The free falling observer would describe these processes
as virtual but the external observer would interpret them as real ones
because of the huge time dilation he experiences with respect  to the
free falling one. If he admits that the trajectory of the shell is
defined only up to a Planck size uncertainties, he would for
sufficiently large retarded time depict as above   the shell as
imbedded in a   polarisation cloud formed of  quanta  at the local
temperature Eq.(27)  superimposed on some, presumably negative, vacuum
Schwartzschild energy. There would be no contradiction with the
description of the free falling observer as long as the mass is
carried away by the Hawking flux is neglected. However the radiation,
in agreement with the thermodynamic description, escapes the
centrifugal barrier through relatively  rare low angular momentum
waves in equilibrium with the thermal cloud.   But if the cloud  does
not separate causally from the collapsing shell by sticking outside
the global event horizon, the description of the external observer
contradicts the image of   the free falling one for whom the classical
shell still carries the total initial mass when it crosses the
horizon. This separation   depicted  in Fig.2 avoids the contradiction
and leads to the bizarre history emerging at the semi-classical level
and the concomitant unitary problem. In our approach the cloud and the
shell do not causally separate even when evaporation occurs and the
full history occurs for external observers in the trivial topological
geometry shown in Fig.3. The history implied by our assumption is
qualitatively the same for a more realistic collapsing star:  after a
short ignition period during which the collapsing star emits thermal
quanta uncorrelated to its structure, the collapse get frozen in
space-time and immersed in a hot   Planckian cloud. There is no a
priori reason which would prevent interactions between the matter
degree of freedom  of the star and the planckian constituents. This
achronon configuration burns away  its mass and structure as any hot
body does within a topologically  trivial background devoid of
horizons and of singularities. Although we have not implemented our
assumption with a  detailed dynamical mechanism, we can therefore see
no reason why the   evaporation process should not be viewed as a
unitary process within this background. The entanglement entropy
responsible for the thermal radiation in the conventional derivation
would be transmuted into more mundane coarse grained entropy stemming
from averaging over quantum phases$^{[6]}$. The price to pay, as in
references [5] and [6] is the apparent discrepancy between the
description of the same event as viewed from free falling or from
static states of motion.

In the achronon background, quantum physics would then apply in the
way we know within our universe at our scale. The consistency of the
overall picture with quantum theory is more problematic. A first
question is whether or not achronon and black hole space-times can
both be contained in the Hilbert space of quantum gravity. As
mentioned above, this appears to be the case in the WKB limit of
quantized general relativity. In fact this was in a previous work the
basic argument in favour of remnants$^{[12],[13]}$. The reason was
that if one takes the usual point of view that a collapsing star
evolves to a classical black hole, the existence of the achronon in
the space of quantum states necessarily leads in this WKB limit to an
infinite degeneracy of the quantum black hole due to the infinite many
ways an achronon can be constructed. This degeneracy implied remnants.
In the present approach the situation is reversed: what is formed is
not a black hole but an achronon and subsequent evaporation does not
pose a degeneracy problem. Every achronon decays in a slightly
different way hidden in phases averaged out in the thermal description
of the radiation.

If indeed both geometries pertain to the same Hilbert space it is
conceivable that           external observer select   achronon
space-times while upon crossing the horizon   freefalling observers
selects enlarged black-hole space-times. This could put  the whole
picture consistently in the framework of quantum physics. Unitarity of
the evaporation process would be conditional to its description within
the achronon background and   transition between different background
would still prevent a exact operational definition of a full unitary
process. This would however be a very weak effect, of the order of the
tunneling probability, namely $\exp (-A/4)$. Maybe third quantization
could allow a general definition of a unitary process in the spaces of
universes\foot{For a review on this subject, see for instance
reference [24]} but without further restrictions  this would make
quantum gravity closer to a metatheory than to a genuine predictive
theory.

To gain further understanding more dynamical ingredients are needed.
Our analysis was phenomenological in character. We have postulated a
stability criterion leading to tamed vacuum fluctuations but we have
not uncovered a dynamical mechanism which does it.  An attempt along
these lines resides perhaps in the type of Hagedorn phase transition
occuring in string theories: transplanckian quanta would condense in
macroscopic strings surrounding the horizon$^{[25],[10]}$. However
such a transition alone is not sufficient to inhibit the fluctuations.
The condensed string should merge with the quantum degrees of the star
itself and thus penetrate inside the star. Otherwise the past history
of the condensate would experience a further blueshift upon entering
the star and rebuild transplanckian effects. It is not clear that
string theory as it stands  contains the required elements to cope
with this problem. It may then be useful to check   the logical
consistency of our assumption by investigating other problems where it
may turn out to be relevant, or self contradictory. An  interesting
problem in this context is offered by primordial  inflation$^{[26]}$.
A permanent de Sitter space time has global horizons which are
different for two distinct comoving motions. Must  the quantum
description be restricted to a single observer or should one work with
a global vacuum which in this case is not marred by a singularity? An
interesting additional problem would arise if de Sitter space is
unstable and evolves, say, into a Robertson-Friedmann universe. This
would get rid of global event horizons but not necessarily of all the
transplanckian fluctuations generated during the de Sitter era. What
happens in this case is not only of academic interest as it may
provide the key for the understanding in a fundamental way the origin
of the   temperature of the  cosmological background radiation. We
hope to return to this problem in future work. \bigskip \centerline
{\bf Acknowledgements} \medskip I thank R. Argurio, R. Brout, S.
Massar and R. Parentani for useful  discussions   and  I am
particularly grateful to Ph. Spindel for  detailed clarifications.

 \vfill \eject \centerline{REFERENCES}

\item{[1]} S.W. Hawking, Commun. Math. Phys. {\bf 43} (1975) 199.

\item{[2]} S.W. Hawking, Phys. Rev. {\bf D14} (1976) 2460 .

\item{[3]} Y. Aharonov, A. Casher and S. Nussinov, Phys. Lett. {\bf
B191} (1987) 51.

\item{[4]} G.'t Hooft, Nucl. Phys. {\bf B335} (1990) 138

\item{[5]}  G.'t Hooft, in the Proceedings of the ``International
Conference on Fundamental Aspects of Quantum Theory" (1992), in Honor
of Y. Aharonov's 60th birthday.\hfill \break C.R. Stephens, G.'t Hooft
and B.F.Whiting, ``{\it Black Hole Evaporation
 without Information Loss"}, Preprint THU-93/20; UF-RAP-93-11, (1993).

\item{[6]} L. Susskind, L. Thorlacius and J. Uglum,
 Phys. Rev. {\bf D48}(1993) 3743.

\item{[7]} T. Jacobson, Phys. Rev. {\bf D44} (1991) 1731.

\item{[8]} K. Schoutens, H. Verlinde, E. Verlinde, ``{\it Black Hole
Evaporation and Quantum Gravity}" Preprint CERN-TH.7142/94, PUPT-1441,
(1994).

\item{[9]} S. Massar and R. Parentani, ``{\it From Vacuum Fluctuations
to Radiation:
 Accelerated Detectors and Black Holes}" Preprint ULB-TH 94/02,
gr-qc/9404057 (1994).

\item{[10]} F. Englert, S. Massar and R. Parentani ``{\it Source
Vacuum Fluctuations of Black Hole Radiance}" Preprint ULB-TH 94/03,
gr-qc/9404026. (1994).

\item{[11]} P. Candelas, Phys. Rev. {\bf D2} (1980) 1541.

\item{[12]} A. Casher and F. Englert, Class. Quantum Grav. {\bf 10}
(1993) 2479.

\item{[13]} A. Casher and F. Englert, ``{\it Entropy Generation
 in Quantum Gravity and Black Hole Remnants}", in ``String Theory,
Quantum Gravity and the Unification of the Fundamental Interactions"
Ed. by M. Bianchi, F. Fucito, E. Marinari, A. Sagnotti, World
Scientific (1993),  gr-qc/9404025.

\item{[14]} W.G. Unruh, Phys. Rev. {\bf D14} (1976) 287.

\item{[15]} R. Parentani and R. Brout, Int. J. Mod. Phys. {\bf D1}
(1992) 169.

\item{[16]} Y. Aharonov, D. Albert, A. Casher and L. Vaidman, Phys.
Lett. {\bf A124} (1987) 199. \hfill \break Y. Aharonov and L. Vaidman,
Phys. Rev. {\bf A41} (1990) 11. \hfill \break Y. Aharonov, J. Anandan,
S. Popescu and L. Vaidman, Phys. Rev. Lett. {\bf 64} (1990) 2965.

\item{[17]} P.C.W. Davies, S.A. Fulling and W.G. Unruh, Phys. Rev.
{\bf D13} (1976) 2720.\hfill \break N.D. Birrel and P.C.W. Davies,
``Quantum Field Theory in Curved Space", Cambridge University Press
(1982).

\item{[18]} J.M. Bardeen, Phys. Rev. Letters {\bf 46} (1981) 382.

\item{[19]} R. Parentani and T.Piran, ``{\it The Internal Geometry of
an Evaporating Black Hole}" hep-th/9405007 (1994).

\item{[20]} J.D. Beckenstein, Phys. Rev. {\bf D7} (1973)  2333.

\item{[21]} J.M. Bardeen, B. Carter and S.W. Hawking, Comm. Math.
Phys. {\bf 31} (1973) 161.

\item{[22]} G. Gibbons and S. Hawking, Phys. Rev. {\bf D15} (1977)
2738, 2752.

\item{[23]} T. Jacobson, Phys. Rev. {\bf D48} (1993) 728.

\item{[24]} T. Banks, ``{\it Report on Progress in Wormhole Physics}"
in ``The Gardener of Eden" Ed. by P. Nicoletopoulos and J. Orloff,
Physicalia Magazine, special issue in honour of R. Brout's birthday,
(1990).
 . \item{[25]} L. Susskind, ``Strings, Black Holes and Lorentz
Contraction"   Preprint SU-ITP-93-21, hep-th/9308139 (1993).
\hfill\break L. Susskind, ``Some speculations about Black Hole Entropy
in String Theory" Preprint RU-93-44, hep-th/9309145 (1993).

\item{[26]} R. Brout, F. Englert and E. Gunzig, Ann. Phys. {\bf 115}
(1978) 78; Gen. Rel. Grav. {\bf 10} (1979) 1; \hfill \break A. Casher
and F. Englert, Phys. Lett. {\bf 104B} (1981) 117. \vfill\eject
\noindent {\bf Figure Captions.} \bigskip \noindent Figure 1. Penrose
Diagram of a  Shell Collapsing to a Black Hole in Absence of
Back-reaction.

The shaded region is the space-time available to external observers.
$S$ and $L$ label respectively generic and light-like shells. The
dotted and dashed lines represents the motion of the centers of vacuum
fluctuation wave-packets; the former is a Hawking photon ancestor
and the latter its correlated partner. \bigskip \noindent Figure 2.
Semi-Classical Picture of the History of a Collapsing Shell.

This Penrose diagram shows the separation outside the shell $S$ of the
apparent horizon $H_a$ from the global event horizon $H$ in the
conventional semi-classical back-reaction picture of a full
evaporation process. The region between these curves contains the
negative energy polarization cloud. The merging of the two horizons at
the singularity is strongly affected by quantum effects and the
drawing there is only indicative. \bigskip \bigskip \noindent Figure
3. Collapse of a Shell to an Achronon and Subsequent Evaporation.

The shell $S$ collapses to a radius close to $2M$ and turns into an
achronon $A$ formed from the shell and its surrounding hot planckian
cloud. The achronon then evaporates to zero radius.

\end